\documentclass[10pt,twocolumn,landscape,a4paper]{article}

\begin{document}

\begin{flushright}
CERN--TH/96--36\\
{\bf hep-th/9602067}\\
\end{flushright}

\vspace{1.5cm}

\begin{center}


\baselineskip16pt

{\large {\bf MASSLESS BLACK HOLES AS BLACK DIHOLES AND QUADRUHOLES}}

\vspace{.9cm}

{\large
{\bf Tom\'as Ort\'{\i}n}
\footnote{E-mail address: {\tt tomas@surya20.cern.ch}}\\
\vspace{.4cm}
{\it C.E.R.N.~Theory Division}\\
{\it CH--1211, Gen\`eve 23, Switzerland}\\
}

\vspace{.8cm}


{\bf Abstract}

\end{center}

\begin{quotation}

\small

\baselineskip11pt

Massless black holes can be understood as bound states of a (positive
mass) extreme $a=\sqrt{3}$ black hole and a singular object with
opposite ({\it i.e.}~negative) mass with vanishing ADM (total) mass but
non-vanishing gravitational field.  Supersymmetric balance of forces is
crucial for the existence of this kind of bound states and explains why
the system does not move at the speed of light.  We also explain how
supersymmetry allows for negative mass as long as it is never isolated
but in bound states of total non-negative mass.

\end{quotation}

\vspace{2cm}

\begin{flushleft}
CERN--TH/96--36\\
February $1996$
\end{flushleft}

\newpage

\pagestyle{plain}


String theory, besides elementary strings, describes many interesting
point-like or extended objects with unbroken supersymmetries and of
solitonic nature.  Thanks to supersymmetry, many properties of these
objects can be reliably studied in the framework of the low-energy
supergravity theory.  Extreme black holes (BHs) are particularly
interesting string theory objects and, the most interesting (and
mysterious) amongst them are perhaps the massless ones.  Their existence
was conjectured by Strominger in Ref.~\cite{kn:S} in the context of
type~II string theory duality phase transitions near conifold points.
Massless BH solutions of the low-energy heterotic string effective
action were recently discovered in Ref.~\cite{kn:B} (see also
Refs.~\cite{kn:K,kn:KL1,kn:KL2}).  Some properties of these objects are

\begin{enumerate}

\item Their (asymptotically flat) canonical metric

\begin{equation}
ds^{2}= \left(1-\frac{D^{2}}{r^{2}}\right)^{-\frac{1}{2}}dt^{2}
-\left(1-\frac{D^{2}}{r^{2}}\right)^{\frac{1}{2}}d\vec{x}^{2}\, ,
\label{eq:metric}
\end{equation}

is singular when $r=D$.  The singularity is a curvature singularity and
the area of spheres of radius $r$ goes to zero in that limit.

\item This metric does not seem to be the extreme limit of any
      non-extreme BH metric\footnote{All this really makes the name
      BH quite inappropriate for them but we will stick to it
      for the moment.}.

\item The expansion of the $g_{tt}$  component of the metric far away
      from the singularity, where the gravitational field is weak is

\begin{equation}
g_{tt}= 1 +\frac{D^{2}}{2}\frac{1}{r^{2}}
+\frac{3D^{4}}{8}\frac{1}{r^{4}}
+{\cal O}\left( \frac{1}{r^{6}} \right)\, .
\label{eq:gtt}
\end{equation}

The coefficient of the $\frac{1}{r}$ term is $-2m$, where $m$ is the ADM
mass.  Then, the ADM mass of these objects is zero (hence the adjective
{\it massless}).  In this limit, $g_{tt}\sim 1+2\Phi$, where $\Phi$ is
the Newtonian gravitational potential.  Therefore,

\begin{equation}
\Phi\sim
\frac{D^{2}}{4}\frac{1}{r^{2}}
+\frac{3D^{4}}{16}\frac{1}{r^{4}}\, ,
\label{eq:Phi}
\end{equation}

and has weakly repulsive (instead of attractive) character when acting
on usual test particles \cite{kn:KL1} (in fact,
$\frac{\partial}{\partial r}g_{tt} < 0\, ,\,\,\, r>|D|$).  If the
Newtonian approximation was valid near the singularity, we could
immediately say that the repulsion grows without bound in its
neighbourhood.

\item They do not seem to move at the speed of light. Usual objects
      with zero rest mass moving at the speed of light have positive
      total energy and non-zero  three-momentum but the total energy of
      masless BHs (their ADM mass) is zero. On the other hand, this
      metric does not admit any  light-like Killing vector and we must
      conclude that the  whole ADM four-momentum of these objects
      actually vanishes. It is  then surprising how, with zero total
      energy and momentum, there is something instead of nothing.

\item When they are rightly embedded in a supergravity theory, they have
      half of $N=2$ or $N=4$ supersymmetries unbroken and the
      low-energy solutions describing them are also exact solutions of
      string theory.

\end{enumerate}

In Ref.~\cite{kn:KL1}, it was observed that {\it the repulsive force
that appears at a finite distance from these objects may be interpreted
as a gravitational interaction with its massive core}. This letter
is an investigation into the nature of the ``massive core'' of
``massless BHs'' for which we will propose a model.

We will start by establishing a heuristic analogy between the expansions
Eqs.~(\ref{eq:gtt},\ref{eq:Phi}) and multipole expansions in
electrostatics.  If we were studying the field created by some charge
distribution confined in a region of space and we had the above
expansions for large $r$ we would immediately say that the charge
distribution has no monopole moment, that is: the total charge is zero.
However, this does not mean that there is no charge!  It just means that
there are as many positive as negative charges, but its number cannot be
deduced from the monopole moment alone.  The existence of terms of
higher order in $\frac{1}{r}$ indicates that the number of positive or
negative charges is not zero (that is why we have a nontrivial field).

The analogy ends here, because multipole momenta terms are not
spherically symmetric.  Let us consider\footnote{I am indebted with
Jorge Russo for a helpful discussion in which he proposed this model to
me.}, though, two charge distributions not confined into a region, with
positive and negative charge respectively, spherically symmetric and
concentric, such that the total (finite) charges are equal (but
opposite).  If the fall-off of charge density of both distributions is
different, the net charge density $\rho (r)$ is different from zero
everywhere but its integral over the whole space is zero.  Then, the net
charge contained in a sphere of radius $r$

\begin{equation}
Q(r)=\int_{S^{3}(r)} d^{3}x^{\prime}\ \rho (r^{\prime})\, ,
\end{equation}

\noindent is a function of $r$ that goes to zero when $r$ goes to
infinity.  Applying Gauss' law one gets the following dependence on $r$
for the electric field

\begin{equation}
E(r) \sim Q(r) \frac{1}{r^{2}}\, .
\end{equation}

Now, if, for instance, $Q(r)\sim \frac{1}{r}$, then $E(r) \sim
\frac{1}{r^{3}}$ and the electrostatic potential $\varphi \sim
\frac{1}{r^{2}}$.  The $\frac{1}{r}$ term appears only when $Q(r) \sim
Q_{0} +\ldots$ and there is a net charge in the whole space.

Then, zero total charge and non-triviality of the field are compatible
with spherical symmetry if the charge distribution is spread over the
whole space.  This would imply for our massless BHs that they could be
composed of two concentric ``charge distributions'' with opposite signs
and vanishing total ``charge''.  The ``charge'' of gravity is the energy
and it is carried by the gravitational field itself.  Then, it is not
necessary to have the whole space filled with positive- and
negative-mass matter.  Comparing the potential Eq.~(\ref{eq:Phi}) with
the above electrostatic potential $\varphi$, we would identify the
$-\frac{D^{2}}{4} \frac{1}{r}$ as the {\it effective mass $m(r)$ at a
distance $r$ from the BH}, much in the same spirit of
Ref.~\cite{kn:KL1}.

There are a number of difficulties with the above hypothesis.  First, it
is usually thought that negative and positive masses in interaction are
unstable and always lead to massless objects moving at the speed of
light: opposite masses repel each other, but a negative mass accelerates
in direction contrary to the force and therefore it follows the positive
mass.  This system does have positive energy, though: the interaction
energy between the objects (the rest mass energies would cancel) and it
is different from a massless BH.

On the other hand, if there was another interaction between these
objects such that the resulting force on each of them is zero, there
would be static configurations describing these two objects in
equilibrium.  Rest masses and interaction energies would cancel and we
would have a massless (zero-energy) system at rest.  Its decay into a
masless system moving at the speed of light ({\it i.~e.} with positive
energy) cannot espontaneously occur.  The existence of additional
charges would also explain why there is no annihilation between positive
and negative masses if the additional charges carried by the objects do
not add up to zero.

The next difficulty would be producing the corresponding static metric.
If a no-force condition holds, one can expect supersymmetry.  The fact
that one of the masses is negative is no obstacle for having
supersymmetry as long as the total ADM mass is not\footnote{It is
tempting to identify the different constants that appear in multi-BH
solutions as the different masses of these objects.  There is, though,
no {\it rigorous} way to assign a value to the mass of each individual
BH.  There is only one asymptotically flat region and only one
ADM mass, the total mass, can be rigorously defined.  One can study
initial-data sets describing $N$ non-extreme BHs which are not
in equilibrium and in them there are $N+1$ asymptotic regions and
individual and total ADM masses can be defined (see, for instance,
\cite{kn:O} and references therein).  The masses turn out to be the
mentioned constants plus interaction energy terms.  These terms vanish
for the static multi-BH solutions and then it is {\it physically
reasonable} to identify the constants with the masses.  We will do so
make heuristic reasonings which will be justified by the results.}.  If
the solution has enough unbroken supersymmetries, the solution should be
stable, at least under static perturbations of the metric.

Then, we should look for supersymmetric, extreme, multi-BH
solutions.  The one we are interested in was found in Refs.~\cite{kn:CY}
and further discussed in Refs.~\cite{kn:CT}.  This solution was later
rediscovered in Ref.~\cite{kn:R} in the framework of the theory
described by the following simple action\footnote{Here we follow the
conventions and notation of Ref.~\cite{kn:KO1}.}

\begin{eqnarray}
S & = & \int dx^{4}\ \sqrt{-g} \left\{ -R
-2\left[ (\partial \phi)^{2} +(\partial \sigma)^{2}
+(\partial \rho)^{2}\right] \right.
\nonumber \\
& & \nonumber \\
& &
+{\textstyle\frac{1}{4}} e^{-2\phi}
\left[
e^{-2(\sigma+\rho)} (F^{(1)1})^{2}
+e^{-2(\sigma-\rho)} (F^{(1)2})^{2}
\right.
\nonumber \\
& & \nonumber \\
& &
\left.
\left.
+e^{2(\sigma+\rho)} (F^{(2)}{}_{1})^{2}
+e^{2(\sigma-\rho)} (F^{(2)}{}_{2})^{2} \right] \right\}\, .
\end{eqnarray}

This action is a truncation of the low-energy effective action of the
heterotic string \cite{kn:MS}.  In particular, $\phi$ is the
four-dimensional dilaton.

The solution is given in terms of four independent harmonic functions
$H^{(1)},K^{(1)},H^{(2)},K^{(2)}$ ($\partial_{\underline{i}}
\partial_{\underline{i}} H =\partial_{\underline{i}}
\partial_{\underline{i}} K=0\, ,\,\,\, i=1,2,3$)

\begin{eqnarray}
ds^{2} & = & U^{-\frac{1}{2}} dt^{2} -U^{\frac{1}{2}} d\vec{x}^{2}\, ,
\hspace{1cm} U= H^{(1)}K^{(1)}H^{(2)}K^{(2)}\, ,
\nonumber \\
& & \nonumber \\
e^{-4\phi} & = & \frac{H^{(1)}H^{(2)}}{K^{(1)}K^{(2)}}\, ,
\hspace{1cm}
e^{-4\sigma} =\frac{H^{(1)}K^{(2)}}{H^{(2)}K^{(1)}}\, ,
\hspace{1cm}
e^{-4\rho} =\frac{H^{(1)}K^{(1)}}{H^{(2)}K^{(2)}}\, ,
\nonumber \\
& & \nonumber \\
F^{(a)1}{}_{t\underline{i}} & = &
c^{(a)} \partial_{\underline{i}}\frac{1}{H^{(1)}}\, ,
\hspace{.5cm}
\tilde{F}^{(a)2}{}_{t\underline{i}} =
d^{(a)} \partial_{\underline{i}}\frac{1}{K^{(1)}}\, ,
\hspace{.5cm}
a=1,2\, .
\end{eqnarray}

\noindent where $(c^{(a)})^{2}=(d^{(a)})^{2}=1$ and

\begin{equation}
\tilde{F}^{(1)2} =e^{-2(\phi+\sigma-\rho)}
{}^{\star}F^{(1)2}\, ,
\hspace{1cm}
\tilde{F}^{(2)}{}_{2} =e^{-2(\phi-\sigma+\rho)}
{}^{\star}F^{(2)}{}_{2}\, ,
\end{equation}

\noindent and ${}^{\star}F$ is the Hodge dual of $F$.  Usually, the
$H$'s and $K$'s are chosen to be strictly positive, that is, all the
constants in

\begin{equation}
H^{(a)}=1 +\sum_{n}\frac{q_{n}^{(a)}}{|\vec{x}-\vec{x}_{n}|}\, ,
\hspace{1cm}
K^{(a)}=1 +\sum_{n}\frac{p^{(a)}_{n}}{|\vec{x}-\vec{x}_{n}|}\, ,
\end{equation}

\noindent are non-negative constants to avoid the occurrence of
singularities in the metric, but, for any positive or negative value of
the constants one gets a solution, and solutions with some negative
$q$'s or $p$'s are what we are after.  Bearing this is mind, and
following Ref.~\cite{kn:R}, let us consider, for simplicity, solutions
of the form

\begin{equation}
H^{(a)} = 1 +\frac{q_{a}}{|\vec{x}-\vec{x}_{1}|}\, ,
\hspace{.5cm}
K^{(a)} = 1 +\frac{p_{a}}{|\vec{x}-\vec{x}_{2}|}\, .
\end{equation}

When all the $q$'s and $p$'s but one vanish, the solution is an
$a=\sqrt{3}$ extreme dilaton BH if the non-vanishing constant is
positive.  Then, if several constants are positive, one can consider
that the above solutions describes as many $a=\sqrt{3}$ BHs in
equilibrium.  The ADM mass of the system is $m ={\textstyle\frac{1}{4}}
(q_{1}+p_{1}+q_{2}+p_{2})$ and would be positive.  When the coordinates
of all the BHs coincide one gets $a=1,1/\sqrt{3},0$ extreme
dilaton BHs (depending on how many constants vanish) and
therefore the above solution, and the corresponding extreme dilaton
BHs can be thought of as describing the external field of a
bound state of ``elementary'' $a=\sqrt{3}$ BHs \cite{kn:R}.

If we now allow for negative constants one immediately sees from the
above mass formula that one could get solutions with $m$ zero or
negative.  We are interested in the former.  They can be thought of as
describing usual extreme $a=\sqrt{3}$ dilaton BHs in equilibrium
amongst them and with some other objects with negative mass\footnote{We
stress again that there is no rigorous way of telling what the mass of
each individual object, although, physically, it is clear that there
must be some negative mass.}.

The simplest massless combination $q_{1}=-q_{2}=q,\,\,\, p_{1}=p_{2}=0$
a BH-anti-BH pair or {\it dihole}.  Here it is clear why we have
something instead of nothing with zero energy.  On the other hand, the
Ricci scalar of a single $a=\sqrt{3}$ extreme BH with metric

\begin{equation}
ds^{2}= \left(1+\frac{q}{r}\right)^{-\frac{1}{2}}dt^{2}
-\left(1+\frac{q}{r}\right)^{\frac{1}{2}}d\vec{x}^{2}\, ,
\end{equation}

\noindent is

\begin{equation}
R=\frac{-q^{2}}{2r(q+r)^{3}}\, .
\end{equation}

When $q$ (the mass) is positive, the singularity is at $r=0$.  When
$q$ is negative, the singularity is at $r=|q|$. We have the same
amount of positive mass matter and negative mass matter placed,
concentrically, at different points and the two corresponding ``charge
distributions'' should only cancel at infinity and, therefore all the
arguments given above apply to this case.  We should get a massless,
nontrivial, point-like object with vanishing ADM mass when the two
massive objects are placed at the same point, and, in fact,
substituting the $H$'s into the metric and placing both BHs in
the same point we recover the massless BH metric
(\ref{eq:metric}) with $D=q$!

Following the same reasoning as in Ref.~\cite{kn:R} we would conclude
that the known massless BHs are the effective field of a bound
state of a pair of objects with opposite masses, or a {\it dihole}.

Another simple massless combination is $q_{1}=-q_{2}=q,\,\,\,
p_{1}=-p_{2}=p$. If the two electric charges $\pm q$ are placed at the
same point and the two magnetic charge $\pm p$ are placed together at a
different point, the resulting solution describes two massless
diholes in equilibrium

\begin{equation}
ds^{2}=
\left[\left(1-\frac{q^{2}}{r_{1}^{2}}\right)
\left(1-\frac{p^{2}}{r_{2}^{2}}\right)\right]^{-\frac{1}{2}}dt^{2}
-\left[\left(1-\frac{q^{2}}{r_{1}^{2}}\right)
\left(1-\frac{p^{2}}{r_{2}^{2}}\right)\right]^{\frac{1}{2}}
d\vec{x}^{2}\, .
\label{eq:metric2}
\end{equation}

When the four charges are placed at the same point one gets a
{\it quadruhole}.  If $p=q$, its metric takes a very simple form

\begin{equation}
ds^{2}= \left(1-\frac{q^{2}}{r^{2}}\right)^{-1}dt^{2}
-\left(1-\frac{q^{2}}{r^{2}}\right)d\vec{x}^{2}\, .
\label{eq:metric3}
\end{equation}

Although more massless solutions are possible these are, perhaps the
most interesting ones, at least to prove our point.

In conclusion, we have exhibited massless extreme BHs solutions that can
be considered as bound states of positive and negative-mass objects
satisfying a no-force condition.

It is difficult to avoid identifying these massless BHs with those
which, according to Strominger \cite{kn:S}, become massless when a
type~II string theory compactified on a Calabi-Yau threefold is near a
conifold singularity of the CY moduli space and which can, in some
cases, condensate \cite{kn:GMS}, giving rise to a phase transition.
This has been proposed in Ref.~\cite{kn:KL2}.  However, we have seen
that the massless BHs found in Ref.~\cite{kn:B} are really composite
objects and they do not correspond to one-particle, but to two-particle
states.  It could well be that Strominger's massless BHs are also
two-particle states.  The fact that the $n_{1}=1$ BHs carries minimal
$Z^{1}$ charge may not be an obstacle for this.  The above massless BHs
also carry minimal charges (of more than one $U(1)$ field, but these
still have to be diagonalized under supergravity).  It is also
irresistible to compare black diholes with Cooper pairs in the BCS
theory of superconductivity.  In spite of the many differences the
analogies are very appealing.

We cannot, however, ignore an important issue: how can supersymmetry be
compatible with objects with negative mass?  The ADM mass of a massless
BH is zero, to start with, and there is no problem in admitting that the
composite object could be supersymmetric.  However, the unbroken
supersymmetry of a composite object is the common sector of the unbroken
supersymmetries of its components: the Killing spinor has to satisfy all
the constraints that the presence of each component imposes (see, for
instance, Refs.~\cite{kn:O1,kn:O2}).  That is: the components have to
admit Killing spinors themselves.

Now, there seems to be a problem with those constituents that have
negative mass.  Certainly, they cannot be supersymmetric: supersymmetry
implies a positivity bound on the mass \cite{kn:WO,kn:FSZ}.  However,
they can still admit Killing spinors (whose existence is necessary
\cite{kn:GWG} but not a sufficient condition to have supersymmetry).  In
fact, it is easy to see by direct calculation that the $a=\sqrt{3}$
multi-BH metrics (for instance) always admit Killing spinors for any
choice of the harmonic function $V$ (see, for instance
Ref.~\cite{kn:KO1})\footnote{This may look strange to the reader that
knows that Killing spinor techniques (Nester constructions) are used to
prove the positivity of the mass and more restrictive bounds
\cite{kn:GHu}.  However, in all cases there are additional assumptions
in the form of inequalities that the energy-momentum tensor has to
satisfy.  They are probably violated in the cases of negative mass.}

At the level of the supersymmetry algebra, the existence of Killing
spinors means that certain supersymmetry charges annihilate the state.
What does this mean for negative mass states?  For an appropriate choice
of the supersymmetry basis, the $N$ extended supersymmetry algebra can
be written in this way \cite{kn:FSZ}

\begin{equation}
\left\{S_{\alpha (\pm)}^{m}, S_{\beta (\pm)}^{\star n} \right\} =
\delta_{\alpha\beta}\ \delta^{mn}\ (m \pm|z_{i}|)\, ,
\end{equation}

\noindent where $i=1,\ldots,[N/2]$ and all other anticommutators vanish.
Since the operators in the l.h.s.~of these equations are positive, we
have the bounds

\begin{eqnarray}
m-|z_{i}| & \geq & 0\, ,
\label{eq:type1} \\
& & \nonumber \\
m+|z_{i}| & \geq & 0\, .
\label{eq:type2}
\end{eqnarray}

Positive mass supersymmetric objects saturate one of the first bounds
Eq.~(\ref{eq:type1}) and satisfy all the others.  The saturation of one
of the first bounds is associated to the existence of a supersymmetry
charge that annihilates the corresponding state.  That charge is
associated to the Killing spinor.  The rest of the charges act
nontrivially and in a way consistent with the supersymmetry algebra on
the state and their action on it generates (shortened) supermultiplets
\cite{kn:GWG}.

For a negative mass object admitting Killing spinors there {\it must} be
a supersymmetry charge that annihilates the corresponding state.  A
supersymmetry bound of the first type Eq.~(\ref{eq:type1}) is
saturated\footnote{Observe that the quadratic form of the bounds,
$m^{2}-|z_{i}|\geq 0$, which is enough to have an extreme solution of
the equations of motion and Killing spinors, can be satisfied by
negative mass objects.} but all bounds of the secondx type are violated.
Then, there are no other supersymmetry charges to complete the algebra,
one cannot build supermultiplets and the state cannot be said
supersymmetric.

In a bound state with a positive mass supersymmetric object, there can
be compensations in the masses and charges and, if the total mass in not
negative, since both components admit Killing spinors, the composite
object can be supersymmetric.

Supersymmetry forbids the existence of isolated negative-mass objects,
but it does not forbid their existence in non-negative mass bound
states, just as quarks do not exist in isolation at low energies.

\vspace{.5cm}

The author would like to thank Luis \'Alvarez-Gaum\'e, Renata Kallosh
and Jorge Russo for helpful discussions.  He would also like to thank
M.M.~Fern\'andez for her support.


\end{document}